\documentstyle[12pt]{article}

\renewcommand\theequation{\arabic{section}.\arabic{equation}}
\begin{document}
\baselineskip=18pt
\begin{titlepage}
\begin{center}
\large{\bf INTEGRABILITY AND SYMMETRY ALGEBRA ASSOCIATED WITH $N=2$ KP FLOWS}
\end{center}
\vspace{2 cm}
\begin{center}
{\bf Sasanka Ghosh}\footnote{email: sasanka@iitg.ernet.in} and 
{\bf Debojit Sarma}\footnote{email: debojit@iitg.ernet.in}\\
{\it Department of Physics, Indian Institute of Technology,}\\
{\it North Guwahati, Guwahati 781 031, INDIA}\\
\end{center}
\vspace{1 cm}


\vspace{3 cm}
\begin{abstract}
We show the complete integrability of $N=2$ nonstandard KP flows
establishing the biHamiltonian structures. One of Hamiltonian structures
is shown to be isomorphic to the nonlinear $N=2$ $\hat W_{\infty}$ algebra
with the bosonic sector having  $\hat W_{1+\infty}\oplus \hat W_{\infty}$ 
structure. A consistent free field representation of the super conformal 
algebra is obtained. The bosonic generators are found to be an admixture of 
free fermions and free complex bosons, unlike the linear one. The fermionic 
generators become exponential in free fields, in general.

\end{abstract}
\end{titlepage}
\newpage

\section{Introduction}

 The close relationship between the conformal algebras and the rich symmetry
associated with integrable systems is well understood. The Hamiltonian
structures of the integrable hierarchies have been found to be isomorphic
to the various higher spin conformal algebras at the classical level.
This was realized when it was shown that the $W_{n}$ algebra incorporates 
in its classical limit the Hamiltonian structure of the nonlinear integrable  
systems i.e. the generalized $n^{th}$ KdV hierarchy \cite{1,2,3}. This technique
of obtaining classical conformal algebras through the integrable hierarchies
is indeed a powerful one and its importance was recognized when the existence
of a new higher spin conformal algebra was realized at the classical level \cite{4}.
Prior to this all the known conformal algebras were obtained first in terms of 
free fields via the bootstrap approach and subsequently their relation to 
the symmetries of the integrable hierarchies was obtained.

Generalized KdV hierarchy, in general, is significant in its own right because 
of its rich symmetry structures. But it becomes physically more relevant since
the equations of motion and symmetries of 2D quantum gravity \cite{5} can be
formulated in terms of KdV-like equations. 

The KP hierarchy incorporates all KdV hierarchies \cite{5a} and thus there
exists the distinct possibility that it is the bedrock of 2D quantum gravity 
\cite{6}. It was believed that the large $n$ limit of $W_{n}$ algebra,
namely $W_{1+\infty}$ and $W_{\infty}$ algebras would provide the necessary 
framework in this direction. The naive approach to the large $n$ limit, however,
gives rise to {\it linear} algebras which are truly infinite dimensional 
symmetry algebras containing all the conformal spins \cite{7}, but the fact
that they are linear prevents them from being the right candidates for universal 
algebras, there being no straightforward mechanism which effectively truncates 
the spin content of these algebras and produces the nonlinear features of $W_{n}$. 
A nonlinear realization of the W algebra namely ${\hat W_{\infty}}$ 
algebra was obtained by the bootstrap approach and identified with the
$2^{nd}$ Hamiltonian structure of the KP hierarchy \cite{8,9}. It is a universal $W$
algebra containing all $W_{n}$ algebras. This was obtained by associating the symmetry 
algebra of the $SL(2,R)_k/U(1)$ coset model with the ${\hat W_{\infty}}$ algebra
characterized by the label $k$ and then by showing that the symmetry 
algebra truncates to $W_n$ algebra for $k =-n$ \cite{10}.
 
Subsequently Manin and Radul \cite{11} provided the
supersymmetric extension of the KP hierarchy, and this was based on the
odd parity superLax formulation. But construction of the Hamiltonian structure 
for odd parity Lax operator, following the Drinfeld Sokolov formalism, is not well 
understood yet \cite{12}. Later on, an even parity Lax operator associated with 
supersymmetric KP hierarchy was obtained \cite{13,14} and a supersymmetric
extension of the linear W algebra was realized \cite{15}. The connection of the
$N=2$ super KdV hierarchies with affine Lie algebras was demonstrated by Inami 
and Kanno \cite{15a,16} which is a step forward towards an $N=2$ super analogue
of the Drinfeld Sokolov formulation. This indicates that there ought to be 
consistent $N=2$ superLax formulation of the super KP hierarchy which should 
be Hamiltonian with respect to the super Gelfand Dikii bracket of the second 
kind and also should reduce to the Lax operators considered by Inami and Kanno 
under suitable reduction. Consequently, the existence of a nonlinear realization 
of the super ${\hat W_{\infty}}$ algebra \cite{4} may be shown through
the super KP formulation. Unlike all other known conformal algebras, there is 
no bootstrap approach to find the higher spin extension of $N=2$ superconformal 
algebra, namely the super $\hat W_{\infty}$ algebra through the free field representations.

While the bosonic KP hierarchy and their connection to 
matrix model have been studied extensively, not much is known about the higher 
spin extension of $N=2$ superconformal symmetry. Moreover, conventional 
formulation of the supersymmetric matrix model failed to describe 
nonperturbative superstring theory; it gives nothing but the ordinary matrix 
model \cite{17}.The super KP formalism may throw some light in this direction
even if no super symmetric extension of the matrix model can be formulated.

Renewed interest in the study of $N=2$ and $N=4$ supersymmetry in the context of
quantum gravity and their connections with integrable systems, opened up a
series of studies relating to the symmetry structures of $N=2$ and $N=4$ supersymmetric
integrable models \cite{18}. A major breakthrough in recent times is the nonperturbative
solution of $N=2$ super Yang-Mills equation and their connection with integrable
systems \cite{19}. Investigations have also been made in recent times to obtain
$\tau$ functions for supersymmetric integrable hierarchies \cite{20}. Motivated by these
works, we intend to further explore the super ${\hat W_{\infty}}$ algebra and 
particularly its free field representation, the significance of which cannot be overestimated
since the underlying representation of a conformal field theory is essentially
a free field representation. Moreover it plays a major role in the classification
of various conformal algebras. Further, the quantization of classical symmetry 
algebras becomes straightforward in terms of free fields.

In this paper, we show that nonstandard supersymmetric KP flows following the
Gelfand Dikii method, are biHamiltonian. We further show that one of the
Hamiltonian structures is a candidate for a higher
spin extension of $N=2$ superconformal algebra and has the required number of 
spin fields and the bosonic sector of the algebra has the right structure with two commuting
sets of bosonic generators. We will also obtain the free field representation of the generators
which turns out to be nonlinear. The generators in the fermionic sector are exponential in the
free fields.
It will be apparent that in the bosonic sector the representation of generators are not trivial
extensions of linear representation. However, unlike the linear symmetry algebras, all
the generators cannot be expressed in terms of the free fields in a closed form. A few of
the lower order generators are explicitly written down and an algorithm for constructing higher
order generators will be indicated. This brings us closer to establishing that
super ${\hat W_{\infty}}$ is a universal symmetry containing all finite dimensional 
bosonic and $N=2$ supersymmetric $W$ algebras. 

The organization of the paper is as follows. In section 2 we introduce the $N=2$ super
KP model and obtain its biHamiltonian structures through the Gelfand Dikii method. This
establishes the complete integrability of the system.
In section 3 we show that the second Hamiltonian structure of
$N=2$ super KP hierarchy exhibits the appropriate structure of a super 
${\hat W_{\infty}}$ algebra. In particular, the bosonic sector of this nonlinear
algebra is shown to possess the required  $W_{1+\infty} \oplus W_{\infty}$ 
structure. We obtain a nonlinear free field representation of the bosonic and 
fermionic generators in a suitable basis in section 4. Unlike the linear 
representation of the bosonic generators which was in terms of bilinears of free
fields -- either fermionic or bosonic; here the consistent free field
representation of these algebras comprises an admixture of free fermions and
free complex scalar fields.  Section 5 is the concluding one.

\section{$N=2$ Super KP Hierarchy and biHamiltonian structures}

In an earlier work \cite{4}, it was shown that with an
even parity superLax operator the Hamiltonian structure leading to a 
nonlinear super
$\hat{W}_{\infty}$ algebra becomes local, thus making it a right candidate 
for a universal
symmetry algebra containing all finite dimensional bosonic as well as $N=2$ supersymmetric
algebras. For completeness and future reference we mention the explicit forms of the Hamiltonian
structure and the dynamical equations of the $N=2$ super KP hierarchy.
The super Lax operator of the associated $N=2$ super KP hierarchy is given by
\begin{equation} 
L = D^{2} + \sum^{\infty }_{i=0} u_{i-1}(X) D^{-i}
\label{II1}
\end{equation}
where, $D$ is the superderivative with $D^{2} = d/dx$ and $u_{i-1}(X)$
are superfields in $X = (x,\theta )$ space, $\theta $ being Grassmann odd
coordinate. The grading of $u_{i-1}(X)$ is $|u_{i-1}|= i$ so that
$u_{2i-1}$ are bosonic superfields, whereas $u_{2i}$ are fermionic ones.

The local Poisson bracket algebra among the coefficient fields $u_{i-1}(X)$ can be
obtained following the method of Gelfand and Dikii \cite{4}. This has the
explicit form
\begin{eqnarray}	
&&\left\{u_{j-1}(X),u_{k-1}(Y)\right\}_{2}=\left[-\sum_{m=0}^{j+1}\left[\begin{array}{c}j+1\\m
\end{array}\right](-1)^{j(k+m+1)+[m/2]}u_{j+k-m}D^{m}\right.\nonumber\\
&&+\sum_{m=0}^{k+1}\left[\begin{array}{c}k+1\\m\end{array}\right]
(-1)^{jm+(k+1)(m+1)}D^{m}u_{j+k-m} \nonumber\\
&&+\sum_{m=0}^{j-1}\sum_{l=0}^{k-1}\left(\left[\begin{array}{c}j\\m+1\end{array}\right]
\left[\begin{array}{c}k\\l+1\end{array}\right]
-\left[\begin{array}{c}j-1\\m\end{array}\right]\left[\begin{array}{c}k-1\\l\end{array}\right]
\right)(-1)^{j(m+1)+k+l+[m/2]} \nonumber\\
&&\times u_{j-m-2}D^{m+l+1}u_{k-l-2}\nonumber\\
&&+\sum_{n=0}^{k-1}\sum_{l=0}^{k-n-1}\left[\begin{array}{c}k-n-1\\l\end{array}\right]
(-1)^{j(n+l)+(l+1)(n+k+1)}u_{n-1}D^{l}u_{j+k-n-l-2}\nonumber\\
&&-\sum_{m=0}^{j+k-n-l-1}\sum_{n=0}^{k-1}\sum_{l=0}^{k-n-1}\left[\begin{array}{c}j-1\\m\end{array}\right]
\left[\begin{array}{c}n+l-1\\l\end{array}\right](-1)^{j(m+n+l+k+1)+n(l+1)+[m/2]} \nonumber\\
&&\left.\times u_{j+k-m-l-2}D^{m+l}u_{n-1}\right]\Delta(X-Y)
\label{II2}
\end{eqnarray} 

Notice that the symmetry algebra (\ref{II2}) possesses the following features
of interest.

1. The algebra is antisymmetric and satisfies the Jacobi identity.

2. The lowest subalgebra contains two super fields, namely $u_{-1}(X)$ and $u_0(X)$ and becomes
isomorphic to the classical analogue of the $N=2$ super conformal algebra.

3. The Hamiltonian structure along with conserved quantites \cite{4}
provides a set of dynamical equations of the $N=2$ KP hierarchy consistent with 
the nonstandard flow equation, namely
\begin{equation}
\frac{dL}{dt_{n}}=[L^{n}_{>0},L]
\label{II3}
\end{equation}
where the super Lax operator $L$ is given in (\ref{II1}). In (\ref{II3}) `$>
0$' implies the +ve part of $L^{n}$ without $D^{0}$ term. The significance of
nonstandard flow equation becomes apparent from the explicit forms of the 
following set of dynamical equations. In fact, the nonstandard flows provide 
the nontrivial dynamics to the the lowest superfield $u_{-1}$ which is 
instrumental in making the Poisson bracket structure (\ref{II2}) local.
The evolution equations corresponding to the lowest three time-flows of the 
hierarchy which follow from (\ref{II1},\ref{II2}) are given below for 
completeness and for future reference.
\begin{eqnarray}
&&\frac{du_{i-1}}{dt_{1}}=u^{[2]}_{i-1}\nonumber\\
&&\frac{du_{i-1}}{dt_{2}}=2u^{[2]}_{i+1}+u^{[4]}_{i-1}+2u_{0}u^{[1]}_{i-1}+2u_{-1}u^{[2]}_{i-1}
-2\left[\begin{array}{c}i+1\\1\end{array}\right]u_{i}u^{[1]}_{-1} \nonumber\\
&&-2(1+(-1)^{i})u_{0}u_{i}+2\sum^{i-1}_{m=0}\left[\begin{array}{c}i\\m+1\end{array}\right]
(-1)^{i+1+[-m/2]}u_{i-m-1}u^{[m+1]}_{0} \nonumber\\
&&+2\sum^{i-1}_{m=0}\left[\begin{array}{c}i+1\\m+2\end{array}\right](-1)^{[m/2]
}u_{i-m-1}u^{[m+2]}_{-1}\nonumber\\
&&\frac{du_{i-1}}{dt_{3}}=3u^{[2]}_{i+3}+3u^{[4]}_{i+1}+u^{[6]}_{i-1}
+6u_{-1}u^{[2]}_{i+1}+3u_{-1}u^{[4]}_{i-1} \nonumber\\
&&-3\left[\begin{array}{c}i+3\\1\end{array}\right]u_{i+2}u^{[1]}_{-1}+3\left
[\begin{array}{c}i+3\\2\end{array}\right]u_{i+1}u^{[2]}_{-1}
+3\left[\begin{array}{c}i+3\\3\end{array}\right]u_{i}u^{[3]}_{-1} \nonumber\\
&&-3(1+(-1)^{i})u_{0}u_{i+2}+3u_{0}u^{[1]}_{i+1}-3(-1)^{i}u_{0}u^{[2]}_{i}
+3u_{0}u^{[3]}_{i-1} \nonumber\\
&&+3\left[\begin{array}{c}i+2\\1\end{array}\right](-1)^{i}u_{i+1}u^{[1]}_{0}-3\left[
\begin{array}{c}i+2\\2\end{array} \right](-1)^{i}u_{i}u^{[2]}_{0}
\nonumber\\
&&+3(u_{1}+2u^{[2]}_{-1}+u^{2}_{-1})u^{[2]}_{i-1}
-3\left[\begin{array}{c}i+1\\1\end{array}\right]u_{i}(u_{1}+u^{[2]}_{-1}+u^{2}_
{-1})^{[1]} \nonumber\\ &&+3(u_{2}+2u_{-1}u_{0}+u^{[2]}_{0})u^{[1]}_{i-1}
-3(1+(-1)^{i})(u_{2}+2u_{-1}u_{0}+u^{[2]}_{0})u_{i} \nonumber\\
&&-3\sum^{i-1}_{m=0}\left[\begin{array}{c}i+3\\m+4\end{array}\right](-1)^{[m/2]
}u_{i-m-1}u^{[m+4]}_{-1} \nonumber\\
&&-3\sum^{i-1}_{m=0}\left[\begin{array}{c}i+2\\m+3\end{array}\right](-1)^{i+[-m
/2]}u_{i-m-1}u^{[m+3]}_{0} \nonumber\\
&&+3\sum^{i-1}_{m=0}\left[\begin{array}{c}i+1\\m+2\end{array}\right](-1)^{[m/2]
}u_{i-m-1} (u_{1}+u^{[2]}_{-1}+u^{2}_{-1})^{[m+2]} \nonumber\\
&&+3\sum^{i-1}_{m=0}\left[\begin{array}{c}i\\m+1\end{array}\right](-1)^{i+[-m/2
]}u_{i-m-1} (u_{2}+2u_{-1}u_{0}+u^{[2]}_{0})^{[m+1]}
\label{II4}
\end{eqnarray}
 In order to show the dynamical equations
associated with the nonstandard flows are completely integrable, we show the
existence of another Hamiltonian structure making the $N=2$ supersymmetric KP
hierarchy biHamiltonian. The super Gelfand Dikii bracket of the first kind is
defined by
 \begin{equation} \left\{ F_P(L) , F_Q(L) \right\}_1=-
Tr\left(\left[P,Q\right]L\right)= -Tr\left(\left[L,P\right]Q\right)
\label{II5}
\end{equation}
where $P$ and $Q$ are the auxiliary fields defined as
\begin{equation}
P = \sum_{j=-2}^{\infty }D^j p_j \qquad ; \qquad Q = \sum_{j=-2}^{\infty}D^j q_j
\label{II6}
\end{equation}
with the grading $|p_j|=|q_j|= j$ so that
the linear functional $F_P(L)$ (and similarly $F_Q(L)$) becomes
\begin{equation}
F_P(L) = Tr (LP) = \sum_{i=0}^{\infty } \int dX (-1)^{i+1} u_{i-1}(X)p_{i-1}(X)
\label{II7}
\end{equation}
Consequently the L.H.S. of (\ref{II5}) becomes
\begin{equation}
\left\{ F_P(L) , F_Q(L) \right\} = \sum_{i,j=0}^{\infty } \int dX \int dY (-1)^{i+j}
p_{i-1}(X) \left\{ u_{i-1}(X) , u_{j-1}(Y) \right\} q_{j-1}(Y)
\label{II8}
\end{equation}
Notice that (\ref{II8}) does not involve terms like $p_{-2}$ and $q_{-2}$
since the superfields begin from $u_{-1}(X)$ in the Lax operator (\ref {II1}).
This consistency is ensured by setting the coefficient of the $D$ term in the
commutator $\left[L,P\right]$ to zero. This leads to the constraint
\begin{eqnarray}
p_{-1}&=&\sum_{j=-1}^{\infty}p_{2j+1}^{[2j+2]}+\sum_{i,m=0}^{\infty}(-1)^{i}
u_{i-1}p_{i+2m+1}^{[2m]} \nonumber\\
&&+\sum_{j=0}^{\infty}\sum_{m=0}^{j-1}\left[\begin{array}{c}j\\m\end{array}\right]
(-1)^{j(m+1)}(p_{j}u_{j-m-2})^{[m]}
\label{II9}
\end{eqnarray}
Using the constraint (\ref{II9}) we obtain from (\ref {II8}) the following
Poisson bracket among the superfields
\begin{eqnarray}
&&\left\{u_{i-1}(X),u_{j-1}(Y)\right\}_1=\left[-\delta_{i,0}D^{j+1}+(-1)^{[i/2]}
\delta_{j,0}D^{i+1}\right. \nonumber\\
&&-\sum_{m=0}^{i+j-1}(-1)^{i(j+m+1)+[m/2]}\left[\begin{array}
{c}i-1\\m\end{array}\right]u_{i+j-m-2}D^{m} \nonumber\\
&&\left.-\sum_{m=0}^{i+j-1}(-1)^{m(i+1)+j(m+1)}\left[\begin{array}
{c}j-1\\m\end{array}\right]D^{m}u_{i+j-m-2}\right]\Delta(
X-Y)
\label{II10}
\end{eqnarray}
It is seen however, that the first Hamiltonian structure above does not 
correctly reproduce the equations of motion (\ref{II4}).
This inconsistency arises because the superLax operator considered is not 
a pure differential operator and was observed also in other cases 
involving pseudo-differential operators \cite{8,14}.
This indicates, like previous cases, a modification of Hamilton's 
equation and consequently the Hamiltonian structure is required. If we modify 
Hamilton's equation of motion to
the form given below,
\begin{eqnarray}
&&\frac{du_{i-1}(X)}{dt_n}=\left\{u_{i-1}(X),H_{n+1}\right\}_{1} \nonumber\\
&&-(-1)^{i}\int dY\left\{u_{i-1}(X),u_{-1}(Y)\right\}_{1}sResL^{n}(Y) \nonumber\\
&&-(-1)^{i}\int dY\left\{u_{i-1}(X),u_{0}(Y)\right\}_{1}sRes\left(L^{n}D^{-1}(Y)\right)
\label{II11}
\end{eqnarray}
where
\begin{eqnarray}
sResL^{n}(X)&=&\sum_{j=0}^{\infty}\delta_{j,0}\frac{\delta H_{n+1}}{\delta u_{j-1}(X)} \nonumber\\
&=&\sum_{j=0}^{\infty}\sum_{l=0}^{j-1}\left[\begin{array}{c}j-1\\l+1\end{array}\right]
D^{l-1}u_{j-l-2}(X)\frac{\delta H_{n+1}}{\delta u_{j-1}(X)}
\label{II12}
\end{eqnarray}
and
\begin{eqnarray}
sResL^{n}D^{-1}(X)&=&\sum_{j=0}^{\infty}(-1)^{j}D_{X}^{j}\frac{\delta H_{n+1}}
{\delta u_{j}(X)} \nonumber\\
&=&\sum_{j=0}^{\infty}\sum_{l=0}^{j-1}\left\{(-1)^{j}D^{-2}u_{j-l-2}(X)D_{X}^{l}
+\left[\begin{array}{c}j-1\\l\end{array}\right]\right.  \nonumber\\
&&\left.\times (-1)^{l(j+1)}D_{X}^{l-2}u_{j-l-2}(X)\right\}\frac{\delta H_{n+1}}
{\delta u_{j-1}(X)}\quad\quad ,
\label{II13}
\end{eqnarray}
it reproduces the equations of motion correctly. Substituting (\ref{II12})
and (\ref{II13}) in (\ref{II11}), the equation
of motion can be rewritten in the form
\begin{equation}
\frac{du_{i-1}(X)}{dt_n}=\left\{u_{i-1}(X),H_{n+1}\right\}_{\tilde 1}
\label{II14}
\end{equation}
which eventually leads us to the correct form of the first Hamiltonian 
structure as
\begin{eqnarray}
&&\left\{u_{i-1}(X),u_{j-1}(Y)\right\}_{\tilde 1}= \nonumber\\
&&\left[-\sum_{m=0}^{i+j-1}\left\{\left[\begin{array}{c}i-1\\m\end{array}\right]
(-1)^{i(j+m+1)+[m/2]}u_{i+j-m-2}D^{m}\right.\right.\nonumber\\
&&+\left.\left[\begin{array}{c}j-1\\m\end{array}\right](-1)^{m(i+1)+j(m+1)}D^{m}u_{i+j-m-2}\right\} \nonumber\\
&&+\sum_{m=0}^{i-1}\sum_{l=0}^{j-1}\left\{\left[\begin{array}{c}j-1\\l+1\end{array}\right]
(-1)^{m(i+1)+j+[m/2]}D^{m}u_{i-m-2}D^{l-1}u_{j-l-2}\right.\nonumber\\
&&-\left[\begin{array}{c}i-1\\m+1\end{array}\right](-1)^{[-m/2]}u_{i-m-2}D^{m-1}u_{j-l-2}D^{l}\nonumber\\
&&+\left[\begin{array}{c}i-1\\m\end{array}\right]\left[\begin{array}{c}j-1\\l+1\end{array}\right]
(-1)^{j+i(m+1)+[m/2]}u_{i-m-2}D^{m+l-1}u_{j-l-2} \nonumber\\
&&+\left[\begin{array}{c}i-1\\m+1\end{array}\right]\left[\begin{array}{c}j-1\\l\end{array}\right]
(-1)^{(l+1)(j+1)+[-m/2]}\nonumber\\
&&\left.\left.\times u_{i-m-2}D^{m+l-1}u_{j-l-2}\right\}\right]\Delta(X-Y)
\label{II15}
\end{eqnarray}
which provides the dynamical equations of the hierarchy associated with the 
super Lax operator (\ref{II1}) and the nonstandard flows (\ref{II3}). It is
evident that the above Hamiltonian structure is manifestly antisymmetric and
satisfies the Jacobi identity. But this Hamiltonian structure, unlike the
earlier one (\ref{II2}), is nonlocal and may not be associated with the
conformal symmetry. This feature is noticed in other supersymmetric integrable
hierarchies also that one of the two Hamiltonian structures becomes nonlocal \cite{14}.
The existence of two Hamiltonian structures, nonetheless, confirms the complete
integrability of the even parity super KP hierarchy.

\setcounter{equation}{0}
\section{Nonlinear Super $\hat W_{\infty}$ Algebra}

In this section we show that the local superalgebra (\ref{II2}) obtained in the
previous section is a higher spin extension of $N=2$ conformal algebra 
containing all conformal spins. The nonlinear nature of superalgebra endows it
with rich algebraic structures.

If the super fields are expressed in component form as
\begin{eqnarray}
&&u_{2i-1}(X)=u^{b}_{2i-1}(x)+\theta u^{f}_{2i-1}(x), \nonumber\\
&&u_{2i}(X)=u^{f}_{2i}(x)+\theta u^{b}_{2i}(x)
\label{III1}
\end{eqnarray}
the odd bosonic fields $u^{b}_{2i-1}(x)$ have conformal weights $i+1$,
whereas the even bosonic fields $u^{b}_{2i}(x)$ have conformal weights
$i+2$ ($i=0,1,2,\ldots $), with respect to the stress tensor
\begin{equation}
T(x)=u^{b}_{0}(x)-\frac{1}{2}\partial_x u^{b}_{-1}(x)
\label{III2}
\end{equation}
On the other hand, both odd fermionic fields $u^{f}_{2i-1}(x)$ and even fermionic
fields $u^{f}_{2i}(x)$ have conformal weights $i+\frac{3}{2}$ ($i=0,1,2,\ldots
$) with respect to the same stress tensor (\ref{III2}). The stress tensor
$T(x)$ belongs to the $N=2$ conformal algebra being a subalgebra of
(\ref{II2}). The conformal weights of the component fields are evident from the
following relations.
\begin{eqnarray}
&&\{T(x),u^{b}_{2i-1}(y)\}=\left[(i+1)u^{b}_{2i-1}(y)\partial_{y}+(u^{b}_{2i-1}(y))'\right. \nonumber\\
&&-\sum^{i-2}_{m=0}(-1)^{m}\left(\begin{array}{c}i\\m+2\end{array}\right)u^{b}_{2i-2m-3}(y)\partial^{m+2}_{y} \nonumber \\
&&\left.-\frac{1}{2}\sum^{i-1}_{m=0}(-1)^{m}\left(\begin{array}{c}i\\m+1\end{array}\right)u^{b}_{2i-2m-3}(y)
\partial^{m+2}_{y}\right]\delta(x-y)\nonumber \\
&&\{T(x),u^{b}_{2i}(y)\}=\left[(i+2)u^{b}_{2i}(y)\partial_{y}+(u^{b}_{2i}(y))'\right. \nonumber\\
&&-\sum^{i-2}_{m=0}(-1)^{m}\left(\begin{array}{c}i+1\\m+2\end{array}\right)u^{b}_{2i-2m-2}(y)\partial^{m+2}_{y} \nonumber \\
&&\left.+\frac{1}{2}\sum^{i-1}_{m=0}(-1)^{m}\left(\begin{array}{c}i+1\\m+1\end{array}\right)u^{b}_{2i-2m-1}(y)
\partial^{m+2}_{y}\right]\delta(x-y) \nonumber\\
&&\{T(x),u^{f}_{2i-1}(y)\}=\left[(i+\frac{3}{2})u^{f}_{2i-1}(y)\partial_{y}+(u^{f}_{2i-1}(y))'\right. \nonumber\\
&&-\sum^{i-2}_{m=0}(-1)^{m}\left(\begin{array}{c}i\\m+2\end{array}\right)u^{f}_{2i-2m-3}(y)\partial^{m+2}_{y} \nonumber \\
&&\left.+\frac{1}{2}\sum^{i-1}_{m=0}(-1)^{m}\left(\begin{array}{c}i\\m+1\end{array}\right)u^{f}_{2i-2m-3}(y)
\partial^{m+2}_{y}\right]\delta(x-y) \nonumber\\
&&\{T(x),u^{f}_{2i}(y)\}=\left[(i+\frac{3}{2})u^{f}_{2i}(y)\partial_{y}+(u^{f}_{2i}(y))'\right. \nonumber\\
&&\left.-\sum^{i-2}_{m=0}(-1)^{m}\left(\begin{array}{c}i+1\\m+2\end{array}\right)u^{f}_{2i-2m-2}(y)
\partial^{m+2}_{y}\right]\delta(x-y)
\label{III3}
\end{eqnarray}
This ensures the presence of a nonlinear supersymmetric conformal algebra in
the Hamiltonian structure (\ref{II2}) of the $N=2$ super KP hierarchy. The 
Poisson brackets among all the component fields are given in appendix A in 
a basis which will be defined later. 

The $W_{1+\infty}\oplus W_{\infty}$ structure of the bosonic sector, however, 
is not apparent in our case from the Poisson bracket between two types 
of bosons, $u_{2i-1}^b$ and $u_{2i}^b$. This is in contrast to the other 
supersymmetric algebras. We shall establish that the bosonic sector of this 
algebra, indeed, has the required $W_{1+\infty} \oplus W_{\infty}$
structure \cite{15}. This step is crucial in obtaining the free field
representation of the generators. To carry out this program, a suitable basis
is required. Notice that the odd bosons themselves $u_{2i-1}$ form
a closed algebra (\ref{AI5}) and consequently for the odd bosons the new set of
generators may be constructed from a linear combination of the fields as
considered in \cite{13}, namely
\begin{equation}
\tilde{W}_{n+1}=\frac{2^{-n}n!}{(2n-1)!!}\sum_{l=0}^{n}\left(\begin{array}{c}n\\l\end{array}\right)
\left(\begin{array}{c}n+l\\l\end{array}\right)u_{2l-1}^{b\;(n-l)}
\label{III4}
\end{equation} 
Since the odd bosons form a closed algebra among themselves, the algebra of the new set of 
generators are also closed and constitute the $\hat{W}_{1+\infty}$ algebra.
Using the Poisson bracket amongst the component fields given in Appendix A, we 
obtain following Poisson brackets for the lower order odd boson generators.
\begin{eqnarray}
&&\{\tilde{W}_{1},\tilde{W}_{1}\}=0\nonumber\\
&&\{\tilde{W}_{2},\tilde{W}_{1}\}=\tilde{W}_{1}\partial\delta (x-y)
\nonumber\\
&&\{\tilde{W}_{2},\tilde{W}_{2}\}=\left[2\tilde{W}_{2}\partial+\tilde{W}_{2}'\right]\delta (x-y)
\nonumber\\
&&\{\tilde{W}_{3},\tilde{W}_{1}\}=2\tilde{W}_{2}\partial\delta (x-y)
\nonumber\\
&&\{\tilde{W}_{3},\tilde{W}_{2}\}=\left[3\tilde{W}_{3}\partial+\tilde{W}_{3}'
+\frac{1}{6}\tilde{W}_{1}\partial^{3}\right]\delta (x-y)
\nonumber\\
&&\{\tilde{W}_{3},\tilde{W}_{3}\}=\left[4\tilde{W}_{4}\partial+2\tilde{W}_{4}'+\frac{3}{5}\tilde{W}_{2}''\partial
+2\tilde{W}_{3}\tilde{W}_{1}\partial-\frac{1}{3}\tilde{W}_{1}''\tilde{W}_{1}\partial\right. \nonumber\\
&&-2\tilde{W}_2^{2}\partial+\frac{1}{2}\tilde{W}_1^{2}\partial+\tilde{W}_{2}'\partial^{2} 
+\frac{2}{3}\tilde{W}_{3}\partial^{3}+\frac{2}{15}\tilde{W}_{2}''' \nonumber\\
&&\left.+\tilde{W}_{3}'\tilde{W}_{1}-\frac{1}{6}\tilde{W}_{1}'''\tilde{W}_{1}
-2\tilde{W}_{2}'\tilde{W}_{2}+\tilde{W}_{3}\tilde{W}_{1}'
+\frac{1}{3}\tilde{W}_{1}''\tilde{W}_{1}'\right]\delta (x-y)
\nonumber\\
&&\{\tilde{W}_{4},\tilde{W}_{1}\}=\left[3\tilde{W}_{3}\partial+\frac{1}{10}\tilde{W}_{1}''\partial
-\frac{3}{10}\tilde{W}_{1}'\partial^{2}
+\frac{1}{10}\tilde{W}_{1}\partial^{3}\right]\delta (x-y)\nonumber\\
&&\{\tilde{W}_{4},\tilde{W}_{2}\}=\left[4\tilde{W}_{4}\partial+\tilde{W}_{4}'
+\frac{7}{10}\tilde{W}_{2}\partial^{3}\right]\delta (x-y)
\label{III5}
\end{eqnarray}

In the even boson, $u_{2i}^{b}$ sector, however, the Poisson bracket relation is complex and it
appears that the generators neither form a closed algebra nor do they commute with the
odd bosons. Apparently, therefore, the direct sum structure is not maintained as required for
the the supersymmetric $W$ algebra. This problem can be circumvented and for the even bosons
also a suitable basis with these desirable properties can be obtained. The first step in making
them commute is to redefine the even bosons as a linear combination of odd and even
bosons of equal spin as in \cite{13}.
\begin{equation}
v_{2j}^{b}=u_{2j+1}^{b}+ u_{2j}^{b}
\label{III6}
\end{equation}
and similarly we choose a linear combination of generators for the odd fermions
\begin{equation}
v_{2j-1}^{f}=u_{2j-1}^{f}-u_{2j}^{f}
\label{III7}
\end{equation}
In the linear super $W_{\infty}$ algebra, this is sufficient to ensure commutation between odd 
and even bosons, and thereby establish the $W_{1+\infty} \oplus W_{\infty}$ structure, but 
in this case it is observed that the odd bosons commute with only the lowest spin even
boson generator i.e.
\begin{equation}
\{u_{2j-1}^{b}, v_{0}^{b}\}=0
\label{III8}
\end{equation}
and the Poisson brackets with higher even bosons  $v_{2j}^{b}$ 
$(j\neq 0)$ are nonzero. Interestingly,
nonlinear combinations of bosons and fermions exist which commute with 
all odd bosons. This may be achieved, as the second step, by taking the most general
combinations of the fields of the appropriate conformal spin and the coefficient of the
terms may be determined by allowing them to commute with the odd bosons. This procedure
can be carried out for all even boson generators thereby yielding the mutually commuting
set of generators. The explicit expressions of a few even boson generators are given below.
\begin{eqnarray}	
W_{2}&=&v_{0}^{b} 
\nonumber\\
W_{3}&=&v_{2}^{b}+\frac{1}{2}v_{0}^{b'}+u_{-1}^{b}v_{0}^{b}+u_{0}^{f}v_{-1}^{f} 
\nonumber\\
W_{4}&=&v_{4}^{b}+v_{2}^{b'}+\frac{1}{5}v_{0}^{b''}+2v_{2}^{b}u_{-1}^{b}+u_{2}^{f}v_{-1}^{f} \nonumber\\
&&+u_{0}^{f}v_{1}^{f}+v_{0}^{b'}u_{-1}^{b}+u_{0}^{f}v_{-1}^{f'}+v_{0}^{b}u_{-1}^{b^{2}}+u_{-1}^{b}u_{0}^{f}v_{-1}^{f}
\label{III9}
\end{eqnarray}
and so on.
The distinguishing character of this set is that spin 3 and higher generators are nonlinear.
For the spin 3 generator it is a bilinear combination of bosonic as well as spin $\frac{3}{2}$
generators. For the spin 4 generator, this combination is more involved, having terms trilinear
in the fields. This indicates that for higher spin generators, the basis becomes more and more
nonlinear. But most importantly,these generators are such that the desired  property is
exhibited, namely
\begin{eqnarray}
\{u_{2j-1}^{b}, W_{2}\}&=&0
\nonumber\\
\{u_{2j-1}^{b}, W_{3}\}&=&0
\nonumber\\
\{u_{2j-1}^{b}, W_{4}\}&=&0 
\label{III10}
\end{eqnarray}
i.e. the new set of even boson generators commute with all the odd bosons.
Moreover the $W$ boson generators produce the requisite form of $\hat W_{\infty}$ 
algebra as can be observed from the following Poisson bracket relations.
\begin{eqnarray}
\{W_{2},W_{2}\}&=&\left[-2W_{2}\partial-W_{2}'\right]\delta (x-y)
\nonumber\\
\{W_{3},W_{2}\}&=&\left[-3W_{3}\partial-W_{3}'\right]\delta (x-y)
\nonumber\\
\{W_{3},W_{3}\}&=&\left[-4W_{4}\partial-2W_{2}^{2}\partial-\frac{9}{20}W_{2}''\partial-2W_{4}'-(W_{2}^{2})'\right. \nonumber\\
&&\left.-\frac{1}{10}W_{2}'''-\frac{3}{4}W_{2}'\partial^{2}-\frac{1}{2}W_{2}\partial^{3} \right]\delta (x-y)\nonumber\\
\{W_{4},W_{2}\}&=&\left[-4W_{4}\partial-W_{4}'-\frac{2}{5}W_{2}\partial^{3}\right]\delta (x-y)
\label{III11}
\end{eqnarray}
This algebra is isomorphic to classical analogue of the $\hat{W}_{\infty}$
algebra \cite{10}. The procedure outlined above, although straightforward, becomes extremely
difficult to use in obtaining the still higher spin generators and to show the closure of the
algebra explicitly. To obtain the genenerators of higher spin we employ a different strategy.
The $W_4$ generator, for example, may be obtained straightforwardly from the
$\{W_{3},W_{3}\}$ Poisson bracket algebra (\ref{III11}) by ensuring the closure of the
algebra following the classical analogue of the $\hat{W}_{\infty}$ algebra \cite {10}.
Importantly the $W_4$ generator thus obtained  matches with that of (\ref{III9}), which
commutes with the odd bosons (\ref{III10}) and the form is unique. Similarly, the
explicit form of the $W_{5}$ generator may be obtained from the Poisson bracket
relation, $\{W_{4},W_{3}\}$ by ensuring the closure of the algebra following
the classical analogue of $\hat{W}_{\infty}$ algebra. It is found that the
leading term of the $\{W_{4},W_{3}\}$ algebra becomes $v_6^b$. This, indeed,
ensures the presence of the $W_{5}$ generator in the algebra. The $W_5$
generator exhibits the explicit form
\begin{eqnarray}
W_{5}&=&v^{b}_{6}+\frac{3}{2}v^{b'}_{4}+\frac{9}{14}v^{b''}_{2}+\frac{1}{14}v^{b'''}_{0}+3u_{-1}^{b}v_{4}^{b}
+3u_{-1}^{b}v_{2}^{b'}+3u_{-1}^{b^2}v_{2}^{b} \nonumber\\
&&+\frac{3}{2}u_{-1}^{b^2}v_{0}^{b'}+u_{-1}^{b^3}v_{0}^{b}-u_{-1}^{b'}v_{0}^{b'}+u_{-1}^{b}v_{0}^{b''}
+\frac{1}{7}\left(u_{-1}^{b}v_{0}^{b}\right)''  \nonumber\\
&&+u_{2}^{f}v_{1}^{f}+u_{4}^{f}v_{-1}^{f}+u_{0}^{f}v_{3}^{f}+\frac{3}{2}u_{0}^{f}v_{1}^{f'}-\frac{1}{2}u_{0}^{f'}v_{1}^{f}
+\frac{1}{2}u_{2}^{f'}v_{-1}^{f} \nonumber\\
&&+\frac{3}{2}u_{2}^{f}v_{-1}^{f'}-u_{0}^{f'}v_{-1}^{f'}+u_{0}^{f}v_{1}^{f''}+\frac{1}{7}\left(u_{0}^{f}v_{-1}^{f}\right)''
+\frac{1}{2}u_{-1}^{b}u_{0}^{f'}v_{-1}^{f}
-\frac{1}{2}u_{-1}^{b'}u_{0}^{f}v_{-1}^{f}\nonumber\\
&&+\frac{3}{2}u_{-1}^{b}u_{0}^{f}v_{-1}^{f'}+u_{-1}^{b^2}u_{0}^{f}v_{-1}^{f}+2u_{-1}^{b}u_{2}^{f}v_{-1}^{f}
+2u_{-1}^{b}u_{0}^{f}v_{1}^{f}+u_{1}^{b}u_{0}^{f}v_{-1}^{f}
\label{III12}
\end{eqnarray}
It is found by explicit calculation that the $W_{5}$ generator commutes with the odd
bosons, $u_{2j-1}^{b}$. The spin six generator $W_6$, in a similar way, may be
obtained from the Poisson bracket relation, $\{W_{4},W_{4}\}$, whose leading term turns out to
be $v_8^b$. Thus the first three generators may be obtained by the
bootstrap approach and from spin four onwards all the generators may be
found following the above procedure. This strategy of obtaining bosonic higher spin generators
evidently guarantees the closure of the algebra being isomorphic
to the classical $\hat{W}_{\infty}$ algebra. We have checked upto spin six
generators explicitly. But to obtain the explicit forms of all higher generators
becomes very difficult, although the strategy is quite clear. Moreover, 
this strategy also ensures that in the bosonic limit the super 
$\hat{W}_{\infty}$ reduces to the $\hat{W}_{\infty}$ algebra. In this way,
we may establish the $\hat{W}_{1+\infty} \oplus \hat{W}_{\infty}$ structure 
in the bosonic sector of the super $\hat{W}_{\infty}$ algebra.

In the fermionic sector both the fermions, $u^f_{2i}$ and $v^f_{2i-1}$ form
closed algebras (\ref{AI7},\ref{AI4}) separately. This reveals that a linear
representation 
such as the one in \cite{13} always exists for the fermions. We shall, however,
show the existence a nonlinear basis for both the fermions in conjunction with
the odd bosons. This follows from the observation that the even fermions,
$u_{2i}^{f}$ as well as the odd fermions, $v_{2i-1}^{f}$ and the odd bosons 
also satisfy separately a subalgebra (see appendix A). Nonetheless, it turns out that
nonlinear basis has an interesting consequence. Both the fermions
satisfy identical algebras in the nonlinear basis with an added advantage of 
generating the minimal algebra. 

The new basis for the even fermion generators may be constructed as a suitable
nonlinear combination of even fermions  $u_{2i}^{f}$ and odd bosons 
$u_{2i-1}^{b}$ in the form
\begin{eqnarray}
\tilde{J}_{3/2}&=&-u_{0}^{f}\nonumber\\
\tilde{J}_{5/2}&=&-u_{2}^{f}-u_{0}^{f'}+u_{-1}^{b}u_{0}^{f} \nonumber\\
\tilde{J}_{7/2}&=&-u_{4}^{f}-u_{2}^{f'}-\frac{5}{4}u_{0}^{f''}+u_{-1}^{b}u_{2}^{f}+u_{1}^{b}u_{0}^{f}
+u_{-1}^{b'}u_{0}^{f}-u_{-1}^{b^{2}}u_{0}^{f}
\label{III13}
\end{eqnarray}
Thus the new set of generators become
more and more nonlinear as in
the even boson case. But it is seen that the nonlinear basis for the even 
fermions (\ref{III13}) can be recast
in terms of the bilinears of the generators only as
\begin{eqnarray}
\tilde{J}_{3/2}&=&-u_{0}^{f}\nonumber\\
\tilde{J}_{5/2}&=&-u_{2}^{f}+\tilde{J}_{3/2}'-\tilde{W}_{1}\tilde{J}_{3/2}\nonumber\\
\tilde{J}_{7/2}&=&-u_{4}^{f}+\tilde{J}_{5/2}'+\frac{1}{4}\tilde{J}_{3/2}''
-\tilde{W}_{1}\tilde{J}_{5/2}+2\tilde{W}_{1}\tilde{J}_{3/2}'\nonumber\\
&&-\tilde{W}_{2}\tilde{J}_{3/2}+\frac{1}{2}\tilde{W}_{1}'\tilde{J}_{3/2}
\label{III14}
\end{eqnarray}
This demonstrates that all the even fermions, in general, can be written as  
bilinears in $\tilde{J}$ and $\tilde{W}$ having the form
\begin{eqnarray}
\tilde{J}_{n+3/2}&=&-u_{2n}^{f}+\frac{2^{-n}(n-1)!}{(2n-1)!!}\sum_{l=0}^{n-1}
\left(\begin{array}{c}n\\l\end{array}\right)
\left(\begin{array}{c}n+l+1\\l+1\end{array}\right)\tilde{J}_{l+3/2}^{(n-l)}
\nonumber\\
&&-\sum_{m=0}^{n-1}\sum_{k=0}^{n-m-1}\sum_{l=0}^{m}B_{kl}^{n}C_{ml}^{n}(-1)^{k+l}
\tilde{J}_{k+3/2}^{(n-m-l-1)}\tilde{W}_{l+1}^{m-l}
\label{III15}
\end{eqnarray}
($n=0,1,2,\ldots $), where $B_{kl}^{n}$ and $C_{ml}^{n}$ are the c-number 
coefficients. While the values of $C_{ml}^{n}$ can be easily extracted 
from (\ref{III4}) by 
determining $u^b_{2i-1}$ in terms of $\tilde{W}_{i+1}$, it appears that
there is no straightforward procedure to determine
the explicit expressions of the $B_{kl}^{n}$ for arbitary spins. The closed
algebras among the even fermions and odd bosons strongly corroborates the 
existence of a nonlinear basis for all higher spin generators and thereby 
ensuring the coefficients $B_{kl}^n$ can always be determined for all higher 
spin generators. The even fermions of higher spins may be obtained from 
(\ref{AI1}) and (\ref{AI9}) through
the Poisson brackets of the even fermions $u_{2k}^{f}$ with
the $W_2$ and $\tilde{W}_2$ generators,
\begin{equation}
\{W_{2},u_{2k}^{f}\}=-u_{2k+2}^{f}-\sum_{m=0}^{k}\left(\begin{array}{c}k+1\\m+1
\end{array}\right)\partial^{m+1} u_{2k-2m}^{f}
\label{IIIA}
\end{equation}
and
\begin{eqnarray}
&&\{\tilde{W}_{2},u_{2k}^{f}\}=-u_{2k+2}^{f}-\frac{1}{2}u_{2k}^{f}+\frac{1}{2}u
_{2k}^{f}\partial\nonumber\\
&&+\sum_{l=0}^{k}\left(\begin{array}
{c}k\\l\end{array}\right)u_{0}^{f}\partial^{l} u_{2k-2l-1}^{b}
-\sum_{l=0}^{k}\left(\begin{array}
{c}k\\l\end{array}\right)u_{-1}^{b}\partial^{l} u_{2k-2l}^{f}
\label{IIIB}
\end{eqnarray}
The presence of the $u_{2k+2}^{f}$ term in (\ref{IIIA},\ref{IIIB}) clearly 
confirms that
the next higher spin generator can be generated from the
preceding one. Finally, (\ref{IIIA}) and (\ref{IIIB}) together with the closure of the fermion generators
determine the higher spin generators explicitly.
As a consequence, the closure of the algebra among the even fermions is 
ensured 
in the nonlinear basis. To show the closure property, the algebra among 
a few even fermion generators are given below.
\begin{eqnarray}
&&\{\tilde{J}_{3/2},\tilde{J}_{3/2}\}=0\nonumber\\
&&\{\tilde{J}_{5/2},\tilde{J}_{3/2}\}=0\nonumber\\
&&\{\tilde{J}_{5/2},\tilde{J}_{5/2}\}=2\tilde{J}_{3/2}'\tilde{J}_{3/2}
\delta (x-y)\nonumber\\                                                
&&\{\tilde{J}_{7/2},\tilde{J}_{3/2}\}=-\tilde{J}_{3/2}'\tilde{J}_{3/2}\delta (x-y)\nonumber\\
&&\{\tilde{J}_{7/2},\tilde{J}_{5/2}\}=\left[-3\tilde{J}_{5/2}\tilde{J}_{3/2}'-3\
\tilde{J}_{5/2}\tilde{J}_{3/2}\partial
+2\tilde{J}_{3/2}'\tilde{J}_{3/2}\partial\right]\delta (x-y)\nonumber\\
&&\{\tilde{J}_{7/2},\tilde{J}_{7/2}\}=\left[6\tilde{J}_{5/2}'\tilde{J}_{5/2}-3\tilde{J}_{5/2}'\tilde{J}_{3/2}'
+3\tilde{J}_{5/2}\tilde{J}_{3/2}''+\frac{3}{4}\tilde{J}_{3/2}''\tilde{J}_{3/2}'\right. \nonumber\\
&&\left.-\frac{5}{4}\tilde{J}_{3/2}'''\tilde{J}_{3/2}-\frac{5}{2}\tilde{J}_{3/2}''\tilde{J}_{3/2}\partial
-\frac{5}{2}\tilde{J}_{3/2}'\tilde{J}_{3/2}\partial^{2}\right]\delta (x-y)
\label{III19}
 \end{eqnarray}                              
It is interesting to note that unlike the linear algebra, the generators do not commute among themselves. This is a
significant departure from the linear representation and compels one to make the free field
representations nonlinear as will be observed later. This difference becomes 
evident from the spin 5/2 generator onwards. For example, the Poisson
bracket of the $\tilde{J}_{5/2}$ generator with itself becomes nonzero. In 
fact, the self brackets of the generators, in general, cannot be made to 
zero by any change of basis. Thus the nonlinear algebra cannot be reduced 
trivially to the linear one. The algebra of $\tilde W$ and $W$ bosons with
even fermions are given in appendix B.

The arguments regarding the existence of the nonlinear basis for the even 
fermion generators
stated above are also valid for the odd fermions since the odd fermions
$v_{2i-1}^{f}$ and  the odd bosons $u_{2i-1}^{b}$
form a closed algebra among themselves (\ref{AI3},\ref{AI4},\ref{AI5}). We can therefore
generate a nolinear basis for the odd fermions $v_{2i-1}^{f}$ as well in terms of odd fermions
and the odd bosons $u_{2i-1}^{b}$. For example a consistent nonlinear
basis for a few odd fermions may be identified as
\begin{eqnarray}
J_{3/2}&=&v_{-1}^{f}\nonumber\\
J_{5/2}&=&-v_{1}^{f}+u_{-1}^{b}v_{-1}^{f}\nonumber\\
J_{7/2}&=&v_{3}^{f}+v_{1}^{f'}+\frac{5}{4}v_{-1}^{f''}-u_{-1}^{b}v_{1}^{f}-u_{1}^{b}v_{-1}^{f}-u_{-1}^{b}v_{-1}^{f'}
+u_{-1}^{b^{2}}v_{-1}^{f}
\label{III20}
\end{eqnarray}
It is straightforward to observe that
the generators in (\ref{III20}) can be rewritten as bilinear combinations
of $\tilde W$ and $J$ like the even fermions. This can also be argued from 
the fact that odd fermions and odd bosons form a subalgebra and thus can be 
generalised for other higher spin generators. The higher spin generators, 
however, may be obtained from the lower ones following the Poisson bracket
relations resulting from (\ref{AI3},\ref{AI8})
\begin{equation}
\{W_{2},v_{2k-1}^{f}\}=v_{2k+1}^{f}-v_{2k-1}^{f}\partial+u_{-1}^{b}v_{2k-1}^{f}
-u_{2k-1}^{b}v_{-1}^{f}
\label{IIIC}
\end{equation}
\begin{eqnarray}
&&\{\tilde{W}_{2},v_{2k-1}^{f}\}=v_{2k+1}^{f}
-\sum_{l=0}^{k-1}\left(\begin{array}{c}k-1\\l\end{array}\right)
u_{-1}^{b}\partial^{l} u_{2k-2l-3}^{f}\nonumber\\
&&+\left[\sum_{m=0}^{k-1}\left(\begin{array}
{c}k\\m+1\end{array}\right)+\frac{1}{2}\sum_{m=0}^{k}\left(\begin{array}
{c}k\\m\end{array}\right)\right]\partial^{m+1}v_{2k-2m-1}^{f}
\label{IIID}
\end{eqnarray}
and by demanding the closure of the algebra among the odd fermions. To 
demonstrate the identical algebra among the odd fermions and the even 
fermions in the nonlinear basis, we give below the algebra among a few odd 
fermion generators explicitly.
\begin{eqnarray}
&&\{J_{3/2},J_{3/2}\}=0 \nonumber\\
&&\{J_{5/2},J_{3/2}\}=0 \nonumber\\
&&\{J_{5/2},J_{5/2}\}=2J_{3/2}'J_{3/2}\delta (x-y) \nonumber\\
&&\{J_{7/2},J_{3/2}\}=-J_{3/2}'J_{3/2}\delta (x-y) \nonumber\\
&&\{J_{7/2},J_{5/2}\}=\left[-3J_{5/2}J_{3/2}'-3J_{5/2}J_{3/2}\partial
+2J_{3/2}'J_{3/2}\partial\right]\delta (x-y)\nonumber\\
&&\{J_{7/2},J_{7/2}\}=\left[6J_{5/2}'J_{5/2}-3J_{5/2}'J_{3/2}'+3J_{5/2}J_{3/2}''+\frac{3}{4} J_{3/2}''J_{3/2}'\right. \nonumber\\
&&\left.-\frac{5}{4} J_{3/2}'''J_{3/2}-\frac{5}{2} J_{3/2}''J_{3/2}\partial-\frac{5}{2} J_{3/2}'J_{3/2}\partial^{2}\right]\delta (x-y)
\label{III21}
\end{eqnarray}
The algebra of odd fermions with even fermions as well as with the bosons are
also given in appendix B.

The super $\hat W_{\infty}$ algebra, in its own right, deserves to be a 
candidate for a universal algebra, unifying all finite dimensional bosonic 
$W$ algebras as well as supersymmetric $W$ algebras. The presence of 
classical analogue of $\hat W_{\infty}$ algebra and a direct sum basis in 
the bosonic sector guarantees that all finite dimensional bosonic $W$ algebras
can be obtained under suitable truncation. Since the super $\hat W_{\infty}$ 
algebra is a higher spin extension of $N=2$ conformal algebra, it is expected 
to contain all finite dimensional $N=2$ supersymmetric $W$ algebras, like the 
bosonic univeral algebra. But a systematic analysis of the truncation of 
super $\hat W_{\infty}$ algebra through some non-compact coset model is yet 
to be studied.

\setcounter{equation}{0}

\section{Free-Field Representation}

 In this section we construct a consistent free field representation of the 
generators discussed in the earlier section. We will show that all the 
generators can be represented by the free complex bosons, $\phi(x,t)$ and 
$\bar{\phi}(x,t)$ and free fermions $\psi(x,t)$ and $\psi^*(x,t)$, which satisfy
the following Poisson bracket algebras
\begin{equation}
\{\psi^{*}(x),\psi(y)\}=\delta (x-y)
\label{IV1}
\end{equation}
and
\begin{equation}
\{\partial\phi (x),\partial\bar{\phi}(y)\}=\partial_{x}\delta (x-y)
\label{IV2}
\end{equation}
The nontrivial Poisson bracket algebras among the even fermions
$\tilde{J}_{n+3/2}$ (\ref{III13},\ref{III14},\ref{III15}) as well as odd
fermions ${J}_{n+3/2}$ (\ref{III20}) ensure a significant departure of the
free field  representations from the linear ones \cite{15}. We will see that the 
representations of the fermion generators become not only nonlinear, and in general
exponential, but also a suitable combinations of both types of bosons and fermions.
This makes the free field representation distinct and important. In the
bosonic sector 
such a nontrivial change in the free field representation is not apparent from 
the Poisson bracket algebras of the generators. But it will be observed that 
the nontrivial Poisson bracket algebras of the fermions become responsible for
a nontrivial change in the free field representation of the bosons over the 
linear ones. 

In order to reproduce the the Poisson brackets for the even fermion generators,
the free field representation of all the generators, in general, turns out to 
be exponential of the boson fields. The explicit forms of the representation a 
few even fermions may be given in order to observe the change in the
fermion sector.  
\begin{eqnarray}
\tilde{J}_{3/2}&=&\psi^{*}\partial\phi e^{i\epsilon\phi_{2}}
\label{IV3a}\\
\tilde{J}_{5/2}&=&[\psi^{*'}\partial\phi+\psi^{*}(a\partial\phi+b\partial\bar{\phi})\partial\phi
-b\psi\psi^{*}\psi^{*'}]e^{i\epsilon\phi_{2}}
\label{IV3b}\\
\tilde{J}_{7/2}&=&\frac{5}{4}(\psi^{*}\partial\phi e^{i\epsilon\phi_{2}})''-\psi^{*'}
(\partial\phi e^{i\epsilon\phi_{2}})'
+b\psi\psi^{*}\psi^{*'}(e^{i\epsilon\phi_{2}})' \nonumber\\
&&+\psi^{*'}(a\partial\phi+b\partial\bar{\phi})\partial\phi e^{i\epsilon\phi_{2}
}
-\psi^{*}(a\partial\phi+b\partial\bar{\phi})(\partial\phi e^{i\epsilon\phi_{2}})
' \nonumber\\
&&+\psi^{*}(a\partial\phi+b\partial\bar{\phi})^{2}\partial\phi e^{i\epsilon\phi_
{2}}
+b\psi^{'}\psi^{*}\psi^{*'}e^{i\epsilon\phi_{2}} \nonumber\\
&&+2b^{2}\psi\psi^{*}\psi^{*'}\partial\bar{\phi} e^{i\epsilon\phi_{2}}
+3ab\psi\psi^{*}\psi^{*'}\partial\phi e^{i\epsilon\phi_{2}}
\label{IV3c}                              
\end{eqnarray}
where, $a$, $b$ and $\epsilon$ are real parameters and 
$\phi_2=\frac{1}{2i}(\phi - \bar{\phi})$. Notice that the significant change 
of the free field representation becomes obvious from $\tilde{J}_{5/2}$ onwards.
This is due to the presence of both $\psi$ and $\psi^*$ fields in the 
representations, the presence of both the fields being essential to 
reproduce the algebra (\ref{III19}). In fact, the presence of both kinds of fermions,
$\psi$ and $\psi^*$ is inevitable to reproduce the nonzero algebras
consistently in the even fermion sector. An algorithmic procedure can be
developed following (\ref{IIIA},\ref{IIIB}) to reproduce the free field representation of all
even fermion generators, but it involves explicit representations of the
$W_2$ and $\tilde W_2$ generators, which will be obtained later on. It is
important to note that in order to  reproduce the even fermion algebras
 (\ref{III19}), the consistency condition demands  that all the parameters are not
independent, but are related by 
\begin{equation}
\frac{b\epsilon}{\sqrt2}-\frac{a\epsilon}{\sqrt2}-ab=1
\label{IV4}
\end{equation}
Evidently, two more parameters still remain arbitrary, which cannot be fixed at 
the classical level. The quantization of the algebra may fix the arbitrary 
parameters through its central charge. The relation (\ref{IV4}) dictates that
all the parameters $a$, $b$ and $\epsilon$ cannot be set to be zero 
simultaneously. This implies the non-linear free field representation cannot be 
reduced to the linear one \cite{15} trivially, which is a crucial observation and will 
also be seen in the odd fermion sector. 

The similar Poisson bracket structures of the odd fermions $J_{n+\frac{3}{2}}$ 
and the even fermions $\tilde{J}_{n+\frac{3}{2}}$, however, indicate the similar 
free field representations of the generators for the odd fermions. We give 
below the representations of the odd fermions upto spin $\frac{7}{2}$. The other
higher spin generators can be constructed in a similar fashion as in the even 
fermion case. The representations of the odd fermions may given as            
\begin{eqnarray}
J_{3/2}&=&-\psi\partial\bar{\phi}e^{-i\epsilon\phi_{2}}
\label{IV5a}\\
J_{5/2}&=&[-\psi^{'}\partial\bar{\phi}+\psi(a\partial\phi+b\partial\bar{\phi})\partial\bar{\phi}
+a\psi\psi^{*}\psi^{'}]e^{-i\epsilon\phi_{2}}
\label{IV5b}\\
J_{7/2}&=&-\frac{5}{4}(\psi\partial\bar{\phi} e^{-i\epsilon\phi_{2}})''+\psi^{'}(\partial\bar{\phi} e^{-i\epsilon\phi_{2}})'
-a\psi\psi^{*}\psi^{'}(e^{-i\epsilon\phi_{2}})' \nonumber\\
&&-\psi^{'}(a\partial\phi+b\partial\bar{\phi})\partial\bar{\phi} e^{-i\epsilon\phi_{2}}
-\psi(a\partial\phi+b\partial\bar{\phi})(\partial\bar{\phi} e^{-i\epsilon\phi_{2}})' \nonumber\\
&&-\psi(a\partial\phi+b\partial\bar{\phi})^{2}\partial\bar{\phi} e^{-i\epsilon\phi_{2}}
-a\psi\psi^{*'}\psi^{'}e^{-i\epsilon\phi_{2}} \nonumber\\
&&+2a^{2}\psi\psi^{*}\psi^{'}\partial\phi e^{-i\epsilon\phi_{2}}
-3ab\psi\psi^{*}\psi^{'}\partial\bar{\phi}e^{-i\epsilon\phi_{2}}
\label{IV5c}
\end{eqnarray}                             
The complex nature of the representations of the odd and even fermions ensure
that the odd and fermions generators cannot be represented by a pair of free 
fields and their complex conjugates, like the linear one. The nonzero values of 
$a$ and $b$ make the
representations different from each other. The representations of other
higher spin generators may follow from (\ref{IIIC},\ref{IIID}) like the even fermion case.

In the bosonic sector, the free field representations become more involved. To be precise, 
in the linear representation \cite{15}, the $W_{1+\infty}$ algebra was realized in terms
of the bilinears of a free fermion field and its cojugate whereas the
$W_{\infty}$  algebra was realized from the bilinears of a free complex scalar
field and its conjugate. This had the advantage of automatically ensuring
the direct sum basis of the generators of these algebras. But such a simple
representation cannot be considered in the present case. This, in turn, leads 
to an inconsistency in the fermion sectors and consequently demands for a 
significant change in the free field representations in the bosonic sector.   

For the odd bosons we have the following consistent representation. The linear
part of the representation may be written in terms of the fermion bilinears.
Thus the lowest one is identical to that of the linear representation, namely
\begin{equation}
\tilde{W}_{1}=-\psi^{*}\psi
\label{IV6}
\end{equation}
On the other hand in spin 2, we have trilinear terms,
\begin{equation}
\tilde{W}_{2}=-\frac{1}{2}(\psi^{*'}\psi -\psi^{*}\psi^{'})-\psi^{*}\psi (a\partial \phi+b\partial \bar{\phi})
\label{IV7}
\end{equation}
and this is the most general form of the spin 2 generator, but involving complex
boson fields. From spin three onwards the representation becomes 
complicated having more and more nonlinear combinations of the free fields. We will 
therefore follow a different strategy from spin three onwards. 
Following a similar
procedure as in the odd boson case, we can obtain the spin 2 generator of the 
other sector.  The most general 
form of the $W_2$ generator is
\begin{equation}
W_{2}=-\partial\phi\partial\bar{\phi}-\psi^{*}\psi (a\partial\phi+b\partial \bar{\phi})
\label{IV8}
\end{equation}
which commutes with odd bosons. For both the spin 2 generators the last term
turns out to be trilinear and more so this is the only possible term that exists
at the spin 2 level being a mixture of bosonic and fermionic fields. 

For higher spin generators, however, the representations of both types of the 
bosonic generators may be obtained from the leading order terms of the Poisson 
brackets, $\{\tilde{J}_{n+3/2},J_{3/2}\}$, the $\tilde{J}_{n+3/2}$ being given
in (\ref{III15}). This will immediately follow from the Poisson bracket relation (\ref{AI2}),
\begin{equation}
\{u_{2n}^{f},v_{-1}^f\} = \left[(u_{2n+1}^b-v_{2n}^b)-\sum_{m=0}^n\left(\begin
{array}{c}{n+1}\\{m+1}\end{array}\right)(-1)^mu_{2n-2m-1}^b\partial^{m+1}\right]
\delta (x-y)
\label{IV9}
\end{equation}
The leading order terms in (\ref {IV9}) evidently will be a combination of
$({\tilde W}_{j+2}-W_{j+2})$ term and the suitable combinations of lower order
spin terms. The representation of $W_{j+2}$ generator and consequently the
${\tilde W}_{j+2}$ generator may be obtained explicitly by exploiting the
commuting property of $W$ and ${\tilde W}$ generators and from the Poisson
bracket relation $\{W_{j+2},W_2\}$. The consistency of these representations
may be checked by comparing the algebra among the other bosonic generators. For 
example, the free field representations ${\tilde W_3}$ and $W_3$ generators may 
be obtained as follows. The Poisson bracket
$\{\tilde{J}_{5/2},J_{3/2}\}$ using the free field representations of the fermion generators
$\tilde{J}_{5/2}$ and $J_{3/2}$  (\ref{IV3b},\ref{IV5a}) is found to be
\begin{eqnarray}
\{\tilde{J}_{5/2},J_{3/2}\}&=&\left[-\partial\phi\partial^{2}\bar{\phi}+\frac{\epsilon}{\sqrt2}\partial\phi\partial\bar{\phi}
(\partial\phi-\partial\bar{\phi})-\partial\phi\partial\bar{\phi}(a\partial\phi+b\partial\bar{\phi})\right.\nonumber\\
&&-\psi^{*'}\psi^{'}-2a\psi^{*}\psi^{'}\partial\phi-b\psi^{*}\psi^{'}\partial\bar{\phi}+\psi^{*'}\psi\partial\bar{\phi}\nonumber\\
&&+a\frac{\epsilon}{\sqrt2}\psi^{*}\psi(\partial\phi)^{2}-b\frac{\epsilon}{\sqrt2}\psi^{*}\psi(\partial\bar{\phi})^{2}
-b\psi^{*}\psi\partial^{2}\bar{\phi}\nonumber\\
&&-\left.\left(\partial\phi\partial\bar{\phi}+\psi^{*'}\psi+2\psi^{*}\psi (a\partial \phi+b\partial \bar{\phi})\right)\partial\right]\delta (x-y)
\label{IV10}
\end{eqnarray}
 comparing the same  with the Poisson bracket of
generators $\tilde{J}_{5/2}$ and $J_{3/2}$ (\ref{AII5b}) allows us to determine
the free field representation of $W_{3}-\tilde{W}_{3}$ from the leading order term
since the free field representation of the non leading order terms are already known. Explicitly,
\begin{eqnarray}
&&W_{3}-\tilde{W}_{3}=\frac{1}{2}[\partial^{2}\phi\partial\bar{\phi}-\partial\phi\partial^{2}
\bar{\phi}]+(-a+\frac{\epsilon}{\sqrt2})(\partial\phi)
^{2}\partial\bar{\phi}-(b+\frac{\epsilon}{\sqrt2})\partial\phi(\partial\bar{\phi})^{2}\nonumber \\
&&+\frac{1}{2}\psi^{*}\psi(a\partial^{2}\phi-b\partial^{2}\bar{\phi})-\psi^{*}
\psi(a\partial\phi+b\partial\bar{\phi})^{2}
+2\partial\phi\partial\bar{\phi}\psi^{*}\psi \nonumber\\
&&+\frac{\epsilon}{\sqrt2}\psi^{*}\psi[a(\partial\phi)^{2}-b(\partial\bar{\phi})
^{2}]-\frac{1}{2}(\psi^{*'}\psi
+\psi^{*}\psi^{'})(a\partial\phi-b\partial\bar{\phi})\nonumber\\
&&+\frac{1}{6}(\psi^{*''}\psi-4\psi^{*'}\psi^{'}+\psi^{*}\psi^{''})+\psi^{*}\psi (a\partial\phi+
b\partial\bar{\phi})^{2} \nonumber\\
&&+(\psi^{*'}\psi-\psi^{*}\psi^{'})(a\partial\phi+b\partial\bar{\phi})
\label{IV11}
\end{eqnarray}
Next, we determine the Poisson bracket between
$W_{3}-\tilde{W}_{3}$ and $W_{2}$. Since the odd and even boson generators
commute (\ref{III10}), this operation allows one to find $W_{3}$ from (\ref{III11})
and hence $\tilde{W}_{3}$. This procedure eventually leads to the following
explicit forms of spin 3 generators
\begin{eqnarray}
\tilde{W}_{3}&=&-\frac{1}{6}(\psi^{*''}\psi-4\psi^{*'}\psi^{'}+\psi^{*}\psi^{''})-\psi^{*}\psi (a\partial\phi+
b\partial\bar{\phi})^{2} \nonumber\\
&&-(\psi^{*'}\psi-\psi^{*}\psi^{'})(a\partial\phi+b\partial\bar{\phi})
\label{IV12}\\
W_{3}&=&\frac{1}{2}[\partial^{2}\phi\partial\bar{\phi}-\partial\phi\partial^{2}
\bar{\phi}]+(-a+\frac{\epsilon}{\sqrt2})(\partial\phi)
^{2}\partial\bar{\phi}-(b+\frac{\epsilon}{\sqrt2})\partial\phi(\partial\bar{\phi})^{2}\nonumber \\
&&+\frac{1}{2}\psi^{*}\psi(a\partial^{2}\phi-b\partial^{2}\bar{\phi})-\psi^{*}
\psi(a\partial\phi+b\partial\bar{\phi})^{2}
+2\partial\phi\partial\bar{\phi}\psi^{*}\psi \nonumber\\
&&+\frac{\epsilon}{\sqrt2}\psi^{*}\psi[a(\partial\phi)^{2}-b(\partial\bar{\phi})
^{2}]-\frac{1}{2}(\psi^{*'}\psi
+\psi^{*}\psi^{'})(a\partial\phi-b\partial\bar{\phi})
\label{IV13}
\end{eqnarray}
where $a$ and $b$ are the same parameters, already introduced in the fermionic
sectors. Notice that the spin 3 generators acquire a complex structure and
possess terms quadrilinear in free fields. In a similar manner
both the spin four generators may be obtained from the Poisson bracket
$\{\tilde{J}_{7/2},J_{3/2}\}$. This constitutes an algorithmic procedure by
means of which free field representations of the higher spin generators may be
constructed. The relation
(\ref{IV4}) dictates the both the parameters $a$ and $b$ cannot be set to be
zero simultaneously, making the representation essentially nonlinear. The
presence of an admixture of the bosonic and fermionic terms, on the
other hand, makes the representation of bosonic sector nontrivial, unlike the
linear case. Importantly, even with the presence of the fermionic and bosonic
fields together, we observe that the odd bosons commute with the even bosons.
All the higher spin generators may be constructed from the algebra amongst the
even and odd fermionic generators (\ref {IV9}) and consistency  of these
representations may be checked by comparing the algebra among the bosonic
generators. But the explicit forms of the higher spin generators in terms of
the free fields become more and more complicated as observed from the spin 3
generators (\ref{IV12},\ref{IV13}). However, the strategy is quite clear.

\setcounter{equation}{0}

\section{Conclusion}

In this paper we have shown that $N=2$ KP hierarchy associated with 
nonstandard flows are biHamiltonian, one of the Hamiltonian structures being
nonlocal. To show the existence of biHamiltonian structures is not
straightforward, as it is an intricate process to obtain the correct Poisson
bracket which makes the $N=2$ KP flows Hamiltonian. Since one of the Hamiltonian structures
is local, it becomes a candidate for a nonlinear super $\hat W_{\infty}$ 
algebra which is
a higher spin extension of $N=2$ superconformal algebra. The bosonic sector 
correctly reproduces the $\hat W_{1+\infty} \oplus \hat W_{\infty}$ structure 
with an appropriate choice of basis. To be explicit, in the even boson sector
the basis becomes highly nontrivial and nonlinear. But we have evoked a novel 
strategy to obtain all the generators. Consequently, the  $\hat W_{\infty}$ 
algebra becomes isomorphic to the classical analogue of the nonlinear 
symmetry considered in \cite{10}. This ensures that the nonlinear super 
$\hat W_{\infty}$ algebra under suitable reduction truncates and 
gives rise to all finite 
dimensional bosonic algebras. In the fermionic sector, the odd and even 
fermions also form closed algebras among themselves in a suitable basis. It
turns out that the algebra among both kinds of fermions becomes distinctly 
different from the linear algebra and more so they form identical algebra 
among themselves. The super $\hat W_{\infty}$ algebra thus deserves to be a 
universal algebra unifying all finite dimensional bosonic as well as 
fermionic $W$ algebras. 
 
The free field representations of the $N=2$ nonlinear super $\hat W_{\infty}$ algebra
are obtained in terms of free complex bosons and fermions. These representations cannot be reduced to the
linear one trivially. This is due the constraint condition (\ref{IV4}).
Moreover, the representation of the bosonic generators in terms of the free fields
possesses a more complex structure having an admixture of complex bosons as well as fermions.
But at the same time the odd and the even bosonic generators mutually commute with each other
maintaining $\hat W_{1+\infty}\oplus \hat W_{\infty}$  structure. This is a nontrivial
generalisation in contrast to the linear representation of the $N=2$ super $\hat W_{\infty}$ algebra.
In the fermionic sector the most general representations become exponential in terms of the free fields.
The free field representations of the $N=2$ nonlinear super $\hat W_{\infty}$ algebra, in fact, is a
major breakthrough in classifying $N=2$ super conformal algebras.

\newpage

\setcounter{equation}{0}

\renewcommand{\theequation}{\Alph{section}\arabic{equation}}

\setcounter{section}{1}

\section*{Appendix A}

Poisson bracket algebra amongst the component fields $u^b_{2i}$, $u^b_{2i-1}$,
$u^f_{2i}$ and $u^f_{2i-1}$.

\begin{eqnarray}
&&\left\{u_{2j-1}^{b}(x),u_{2k}^{f}(y)\right\}=\left[-\sum_{m=0}^{j}\left(\begin{array}{c}j\\m\end{array}\right)
(-1)^{m}u_{2j+2k-2m}^{f}\partial^{m} \right.\nonumber\\
&&+\sum_{m=0}^{j-1}\sum_{l=0}^{k}\left(\begin{array}{c}j-1\\m\end{array}\right)\left(\begin{array}{c}k\\l\end{array}\right)
(-1)^{m}u_{2j-2m-2}^{f}\partial^{m+l}u_{2k-2l-1}^{b} \nonumber\\
&&-\sum_{m=0}^{j-1}\sum_{l=0}^{k-1}\left(\begin{array}{c}j\\m+1\end{array}\right)\left(\begin{array}{c}k\\l+1\end{array}\right)
(-1)^{m}u_{2j-2m-3}^{b}\partial^{m+l+1}u_{2k-2l-2}^{f} \nonumber\\
&&+\sum_{n=0}^{k-1}\sum_{l=0}^{k-n-1}\left(\begin{array}{c}k-n-1\\l\end{array}\right)u_{2n}^{f}\partial^{l}u_{2j+2k-2n-2l-3}^{b} \nonumber\\
&&-\sum_{m=0}^{j+k-n-l-1}\sum_{n=0}^{k}\sum_{l=0}^{k-n}\left(\begin{array}{c}j-1\\m\end{array}\right)
\left(\begin{array}{c}n+l-1\\l\end{array}\right)(-1)^{m} \nonumber\\
&&\left.\times u_{2j+2k-2m-2n-2l-2}^{f}\partial^{m+l}u_{2n-1}^{b}\right]\delta(x-y)
\label{AI1}
\end{eqnarray}

\begin{eqnarray}
&&\left\{u_{2j}^{f}(x),u_{2k}^{f}(y)\right\}= \nonumber\\
&&\left[-\sum_{m=0}^{j-1}\sum_{l=0}^{k-1}\left(\begin{array}{c}j\\m+1\end{array}\right)
\left(\begin{array}{c}k\\l+1\end{array}\right)(-1)^{m}u_{2j-2m-2}^{f}\partial^{m+l+1}u_{2k-2l-2}^{f}\right. \nonumber\\
&&-\sum_{n=0}^{k-1}\sum_{l=0}^{k-n-1}\left(\begin{array}{c}k-n-1\\l\end{array}\right)
u_{2n}^{f}\partial^{l}u_{2j+2k-2n-2l-2}^{f} \nonumber\\
&&+\sum_{m=0}^{j+k-n-l-1}\sum_{n=0}^{k-1}\sum_{l=0}^{k-n-1}\left(\begin{array}{c}j\\m\end{array}\right)
\left(\begin{array}{c}n+l\\l\end{array}\right) \nonumber\\
&&\left.\times (-1)^{m}u_{2j+2k-2m-2n-2l-2}^{f}\partial^{m+l}u_{2n}^{f}\right]\delta(x-y)
\label{AI7}
\end{eqnarray}

\newpage
\begin{eqnarray}
&&\left\{u_{2j-1}^{b}(x),v_{2k-1}^{f}(y)\right\}=\left[\sum_{m=0}^{k}\left(\begin{array}{c}k\\m\end{array}\right)
\partial^{m}v_{2j+2k-2m-1}^{f}\right. \nonumber\\
&&+\sum_{m=0}^{j-1}\sum_{l=0}^{k-1}\left\{\left(\begin{array}{c}j-1\\m\end{array}\right)\left(\begin{array}{c}k-1\\l\end{array}\right)
-\left(\begin{array}{c}j\\m+1\end{array}\right)\left(\begin{array}{c}k\\l+1\end{array}\right)\right\} \nonumber\\
&&\times (-1)^{m}u_{2j-2m-3}^{b}\partial^{m+l+1}v_{2k-2l-3}^{f} \nonumber\\
&&+\sum_{n=0}^{k-1}\sum_{l=0}^{k-n-1}\left(\begin{array}{c}k-n-1\\l\end{array}\right)u_{2n-1}^{b}\partial^{l}v_{2j+2k-2n-2l-3}^{f} \nonumber\\
&&-\sum_{m=0}^{j+k-n-l-1}\sum_{n=0}^{k-1}\sum_{l=0}^{k-n-1}\left(\begin{array}{c}j-1\\m\end{array}\right)
\left(\begin{array}{c}n+l-1\\l\end{array}\right) \nonumber\\
&&\left.\times (-1)^{m}u_{2j+2k-2m-2n-2l-3}^{b}\partial^{m+l}v_{2n-1}^{f}\right]\delta(x-y)
\label{AI3}
\end{eqnarray}

\begin{eqnarray}
&&\left\{v_{2j-1}^{f}(x),v_{2k-1}^{f}(y)\right\}= \nonumber\\
&&\left[\sum_{m=0}^{j-1}\sum_{l=0}^{k-1}\left\{\left(\begin{array}{c}j-1\\m\end{array}\right)
\left(\begin{array}{c}k-1\\l\end{array}\right)-\left(\begin{array}{c}j\\m+1\end{array}\right)
\left(\begin{array}{c}k\\l+1\end{array}\right)\right\}\right. \nonumber\\
&&\times (-1)^{m}v_{2j-2m-3}^{f}\partial^{m+l+1}v_{2k-2l-3}^{f} \nonumber\\
&&+\sum_{n=0}^{k-1}\sum_{l=0}^{k-n-1}\left(\begin{array}{c}k-n-1\\l\end{array}\right)
v_{2n-1}^{f}\partial^{l}v_{2j+2k-2n-2l-3}^{f} \nonumber\\
&&-\sum_{m=0}^{j+k-n-l-1}\sum_{n=0}^{k-1}\sum_{l=0}^{k-n-1}\left(\begin{array}{c}\j-1\\m\end{array}\right)
\left(\begin{array}{c}n+l-1\\l\end{array}\right) \nonumber\\
&&\left.\times (-1)^{m}v_{2j+2k-2m-2n-2l-3}^{f}\partial^{m+l}v_{2n-1}^{f}\right]\delta(x-y) 
\label{AI4}
\end{eqnarray}

\newpage

\begin{eqnarray}
&&\left\{u_{2j-1}^{b}(x),u_{2k-1}^{b}(y)\right\}=\left[-\sum_{m=0}^{j}\left(\begin{array}{c}j\\m\end{array}\right)
(-1)^{m}u_{2j+2k-2m-1}^{b}\partial^{m}\right. \nonumber\\
&&+\sum_{m=0}^{k}\left(\begin{array}{c}k\\m\end{array}\right)\partial^{m}u_{2j+2k-2m-1}^{b} \nonumber\\
&&+\sum_{m=0}^{j-1}\sum_{l=0}^{k-1}\left\{\left(\begin{array}{c}j-1\\m\end{array}\right)
\left(\begin{array}{c}k-1\\l\end{array}\right)-\left(\begin{array}{c}j\\m+1\end{array}\right)
\left(\begin{array}{c}k\\l+1\end{array}\right)\right\} \nonumber\\
&&\times (-1)^{m}u_{2j-2m-3}^{b}\partial^{m+l+1}u_{2k-2l-3}^{b} \nonumber\\
&&+\sum_{n=0}^{k-1}\sum_{l=0}^{k-n-1}\left(\begin{array}{c}k-n-1\\l\end{array}\right)
u_{2n-1}^{b}\partial^{l}u_{2j+2k-2n-2l-3}^{b} \nonumber\\
&&-\sum_{m=0}^{j+k-n-l-1}\sum_{n=0}^{k-1}\sum_{l=0}^{k-n-1}\left(\begin{array}{c}j-1\\m\end{array}\right)
\left(\begin{array}{c}n+l-1\\l\end{array}\right) \nonumber\\
&&\left.\times (-1)^{m}u_{2j+2k-2m-2n-2l-3}^{b}\partial^{m+l}u_{2n-1}^{b}\right]\delta(x-y)
\label{AI5}
\end{eqnarray}

\begin{eqnarray}
&&\left\{v_{2j}^{b}(x),v_{2k}^{b}(y)\right\}= \nonumber\\
&&\left[\sum_{m=0}^{j+1}\left(\begin{array}{c}j+1\\m\end{array}\right)(-1)^{m}v_{2j+2k-2m+2}^{b}
\partial^{m}-\sum_{m=0}^{k+1}\left(\begin{array}{c}k+1\\m\end{array}\right)\partial^{m}v_{2j+2k-2m+2}^{b}\right. \nonumber\\
&&-\sum_{m=0}^{j-1}\sum_{l=0}^{k-1}\left(\begin{array}{c}j\\m+1\end{array}\right)\left(\begin{array}{c}k\\l+1\end{array}\right)
(-1)^{m}v_{2j-2m-2}^{b}\partial^{m+l+1}v_{2k-2l-2}^{b} \nonumber\\
&&-\sum_{l=0}^{k}\left(\begin{array}{c}k\\l\end{array}\right)u_{-1}^{b}\partial^{l}v_{2j+2k-2l}^{b}
-\sum_{l=0}^{k}\left(\begin{array}{c}k\\l\end{array}\right)u_{2j}^{f}\partial^{l}v_{2k-2l-1}^{f} \nonumber\\
&&+\sum_{m=0}^{j}\left(\begin{array}{c}j\\m\end{array}\right)(-1)^{m}v_{2j+2k-2m}^{b}\partial^{m}u_{-1}^{b}
-\sum_{m=0}^{j}\left(\begin{array}{c}j\\m\end{array}\right)(-1)^{m}v_{2j-2m-1}^{f}\partial^{m}u_{2k}^{f} \nonumber\\
&&-\sum_{n=0}^{k-1}\sum_{l=0}^{k-n-1}\left(\begin{array}{c}k-n-1\\l\end{array}\right)v_{2n}^{b}\partial^{l}v_{2j+2k-2n-2l-2}^{b} \nonumber\\
&&+\sum_{m=0}^{j+k-n-l-1}\sum_{n=0}^{k-1}\sum_{l=0}^{k-n-1}\left(\begin{array}{c}j\\m\end{array}\right)
\left(\begin{array}{c}n+l\\l\end{array}\right) \nonumber\\
&&\left.\times (-1)^{m}v_{2j+2k-2m-2n-2l-2}^{b}\partial^{m+l}v_{2n}^{b}\right]\delta(x-y)
\label{AI6}
\end{eqnarray}

\newpage

\begin{eqnarray}
&&\left\{u_{2j}^{f}(x),v_{2k-1}^{f}(y)\right\}= \nonumber\\
&&\left[\sum_{m=0}^{j+1}\left(\begin{array}{c}j+1\\m\end{array}\right)(-1)^{m}u_{2j+2k-2m+1}^{b}\partial^{m}
-\sum_{m=0}^{k}\left(\begin{array}{c}k\\m\end{array}\right)\partial^{m}v_{2j+2k-2m}^{b}\right. \nonumber\\
&&-\sum_{m=0}^{j-1}\sum_{l=0}^{k-1}\left(\begin{array}{c}j\\m+1\end{array}\right)
\left(\begin{array}{c}k\\l+1\end{array}\right)(-1)^{m}u_{2j-2m-2}^{f}\partial^{m+l+1}v_{2k-2l-3}^{f} \nonumber\\
&&-\sum_{m=0}^{j}\left(\begin{array}{c}j\\m\end{array}\right)(-1)^{m}u_{2j-2m-1}^{b}\partial^{m}u_{2k-1}^{b} 
+\sum_{m=0}^{j}\left(\begin{array}{c}j\\m\end{array}\right)(-1)^{m}u_{2j+2k-2m-1}^{b}\partial^{m}u_{-1}^{b} \nonumber\\
&&-\sum_{n=0}^{k-1}\sum_{l=0}^{k-n-1}\left(\begin{array}{c}k-n-1\\l\end{array}\right)
u_{2n-1}^{f}\partial^{l}u_{2j+2k-2n-2l-2}^{b} \nonumber\\
&&+\sum_{m=0}^{j+k-n-l-1}\sum_{n=0}^{k-1}\sum_{l=0}^{k-n-1}\left(\begin{array}{c}j\\m\end{array}\right)
\left(\begin{array}{c}n+l\\l\end{array}\right) \nonumber\\
&&\left.\times (-1)^{m}u_{2j+2k-2m-2n-2l-3}^{b}\partial^{m+l}v_{2n}^{b}\right]\delta(x-y)
\label{AI2}
\end{eqnarray}

\begin{eqnarray}
&&\left\{v_{2j}^{b}(x),v_{2k-1}^{f}(y)\right\}=\left[\sum_{m=0}^{j+1}\left(\begin{array}{c}j+1\\m\end{array}\right)
(-1)^{m}v_{2j+2k-2m+1}^{f}\partial^{m}\right. \nonumber\\
&&-\sum_{m=0}^{j-1}\sum_{l=0}^{k-1}\left(\begin{array}{c}j\\m+1\end{array}\right)\left(\begin{array}{c}k\\l+1\end{array}\right)
(-1)^{m}v_{2j-2m-2}^{b}\partial^{m+l+1}v_{2k-2l-3}^{f} \nonumber\\
&&+\sum_{m=0}^{j}\left(\begin{array}{c}j\\m\end{array}\right)(-1)^{m}v_{2j+2k-2m-1}^{f}\partial^{m}u_{-1}^{b} \nonumber\\
&&-\sum_{m=0}^{j}\left(\begin{array}{c}\j\\m\end{array}\right)(-1)^{m}v_{2j-2m-1}^{f}\partial^{m}u_{2k-1}^{b} \nonumber\\
&&-\sum_{n=0}^{k-1}\sum_{l=0}^{k-n-1}\left(\begin{array}{c}k-n-1\\l\end{array}\right)
v_{2n-1}^{f}\partial^{l}v_{2j+2k-2n-2l-2}^{b} \nonumber\\
&&+\sum_{m=0}^{j+k-n-l-1}\sum_{n=0}^{k-1}\sum_{l=0}^{k-n-1}\left(\begin{array}{c}j\\m\end{array}\right)
\left(\begin{array}{c}n+1\\l\end{array}\right) \nonumber\\
&&\left.\times (-1)^{m}v_{2j+2k-2m-2n-2l-3}^{f}\partial^{m+l}v_{2n}^{b}\right]\delta(x-y)
\label{AI8}
\end{eqnarray}

\newpage

\begin{eqnarray}
&&\left\{v_{2j}^{b}(x),u_{2k}^{f}(y)\right\}=\left[-\sum_{m=0}^{k+1}\left(\begin{array}{c}k+1\\m\end{array}\right)
\partial^{m}u_{2j+2k-2m+2}^{f}\right. \nonumber\\
&&-\sum_{m=0}^{j-1}\sum_{l=0}^{k-1}\left(\begin{array}{c}j\\m+1\end{array}\right)\left(\begin{array}{c}k\\l+1\end{array}\right)
(-1)^{m}v_{2j-2m-2}^{b}\partial^{m+l+1}u_{2k-2l-2}^{f} \nonumber\\
&&+\sum_{l=0}^{k}\left(\begin{array}{c}k\\l\end{array}\right)u_{2j}^{f}\partial^{l}u_{2k-2l-1}^{b}-\sum_{l=0}^{k}
\left(\begin{array}{c}k\\l\end{array}\right)u_{-1}^{b}\partial^{l}u_{2j+2k-2l}^{f} \nonumber\\
&&-\sum_{n=0}^{k-1}\sum_{l=0}^{k-n-1}\left(\begin{array}{c}k-n-1\\l\end{array}\right)v_{2n}^{b}
\partial^{l}u_{2j+2k-2n-2l-2}^{f} \nonumber\\
&&+\sum_{m=0}^{j+k-n-l-1}\sum_{n=0}^{k-1}\sum_{l=0}^{k-n-1}\left(\begin{array}{c}j\\m\end{array}\right)
\left(\begin{array}{c}n+l\\l\end{array}\right) \nonumber\\
&&\left.\times (-1)^{m}v_{2j+2k-2m-2n-2l-2}^{b}\partial^{m+l}u_{2n}^{f}\right]\delta(x-y)
\label{AI9}
\end{eqnarray}

\begin{eqnarray}
&&\left\{u_{2j-1}^{b}(x),v_{2k}^{b}(y)\right\}= \nonumber\\
&&\left[-\sum_{m=0}^{j-1}\sum_{l=0}^{k}\left(\begin{array}{c}j-1\\m\end{array}\right)
\left(\begin{array}{c}k\\l\end{array}\right)(-1)^{m}u_{2j-2m-2}^{f}\partial^{m+l}v_{2k-2l-1}^{f}\right. \nonumber\\
&&-\sum_{m=0}^{j-1}\sum_{l=0}^{k-1}\left(\begin{array}{c}j\\m+1\end{array}\right)\left(\begin{array}{c}k\\l+1\end{array}\right)
(-1)^{m}u_{2j-2m-3}^{b}\partial^{m+l+1}v_{2k-2l-2}^{b} \nonumber\\
&&-\sum_{n=0}^{k-1}\sum_{l=0}^{k-n-1}\left(\begin{array}{c}k-n-1\\l\end{array}\right)u_{2n}^{f}\partial^{l}v_{2j+2k-2n-2l-3}^{f} \nonumber\\
&&+\sum_{m=0}^{j+k-n-l-1}\sum_{n=0}^{k}\sum_{l=0}^{k-n}\left(\begin{array}{c}j-1\\m\end{array}\right)
\left(\begin{array}{c}n+l-1\\l\end{array}\right) \nonumber\\
&&\left.\times (-1)^{m}u_{2j+2k-2m-2n-2l-2}^{f}\partial^{m+l}v_{2n-1}^{f}\right]\delta(x-y)
\label{AI10}
\end{eqnarray}

\newpage

\setcounter{equation}{0}

\setcounter{section}{2} 

\section*{Appendix B}

The Poisson brackets between odd bosons and even fermions are
\begin{eqnarray}
\{\tilde{W}_{1},\tilde{J}_{3/2}\}&=&-\tilde{J}_{3/2}\delta (x-y)
\label{AII1a}\\
\{\tilde{W}_{2},\tilde{J}_{3/2}\}&=&\left[-\tilde{J}_{5/2}+\frac{1}{2}\tilde{J}_{3/2}'+\frac{1}{2}\tilde{J}_{3/2}\partial
-\tilde{W}_{1}\tilde{J}_{3/2}\right]\delta (x-y)
\label{AII1b}\\
\{\tilde{W}_{1},\tilde{J}_{5/2}\}&=&\left[-\tilde{J}_{5/2}+\tilde{J}_{3/2}'+\tilde{J}_{3/2}\partial\right]\delta (x-y)
\label{AII1c}\\
\{\tilde{W}_{2},\tilde{J}_{5/2}\}&=&\left[-\tilde{J}_{7/2}+\frac{1}{2}\tilde{J}_{5/2}'+\frac{3}{4}\tilde{J}_{3/2}''
-\tilde{W}_{1}\tilde{J}_{5/2}\right. \nonumber\\
&&\left.+\frac{3}{2}\tilde{J}_{5/2}\partial-\tilde{J}_{3/2}'\partial-\frac{1}{2}\tilde{J}_{3/2}\partial^{2}\right]\delta (x-y)
\label{AII1d}
\end{eqnarray}
The Poisson brackets between odd bosons and odd fermions are:
\begin{eqnarray}
\{\tilde{W}_{1},J_{3/2}\}&=&J_{3/2}\delta (x-y)
\label{AII2a}\\
\{\tilde{W}_{2},J_{3/2}\}&=&\left[-J_{5/2}+\frac{1}{2}J_{3/2}'+\frac{1}{2}J_{3/2}\partial+\tilde{W}_{1}J_{3/2}\right]\delta (x-y)
\label{AII2b}\\
\{\tilde{W}_{1},J_{5/2}\}&=&\left[J_{5/2}-J_{3/2}'-J_{3/2}\partial\right]\delta (x-y)
\label {AII2c}\\
\{\tilde{W}_{2},J_{5/2}\}&=&\left[-J_{7/2}+\frac{1}{2}J_{5/2}'
+\frac{3}{4}J_{3/2}''+\tilde{W}_{1}J_{5/2}\right. \nonumber\\
&&\left.+\frac{3}{2}J_{5/2}\partial-J_{3/2}'\partial-\frac{1}{2}J_{3/2}\partial^{2}\right]\delta (x-y)
\label{AII2d}
\end{eqnarray}
The Poisson brackets between even bosons and even fermions are:
\begin{eqnarray}
\{W_{2},\tilde{J}_{3/2}\}&=&\left[-\tilde{J}_{5/2}-\tilde{J}_{3/2}\partial-\tilde{W}_{1}\tilde{J}_{3/2}\right]\delta (x-y)
\label{AII3a}\\
\{W_{2},\tilde{J}_{5/2}\}&=&\left[-\tilde{J}_{7/2}-\tilde{J}_{5/2}'
+\frac{5}{4}\tilde{J}_{3/2}''-\tilde{W}_{1}\tilde{J}_{5/2}-
\tilde{J}_{5/2}\partial\right]\delta (x-y)
\label{AII3b}
\end{eqnarray}
The Poisson brackets between even bosons and odd fermions are:
\begin{eqnarray}
\{W_{2},J_{3/2}\}&=&\left[-J_{5/2}-J_{3/2}\partial+\tilde{W}_{1}J_{3/2}\right]\delta (x-y)
\label{AII4a}\\
\{W_{2},J_{5/2}\}&=&\left[-J_{7/2}-J_{5/2}'+\frac{5}{4} J_{3/2}''+\tilde{W}_{1}J_{5/2}-J_{5/2}\partial\right]\delta (x-y)
\label{AII4b}
\end{eqnarray}
The Poisson bracket between even fermions and odd fermions are:
\begin{eqnarray}
&&\{\tilde{J}_{3/2},J_{3/2}\}=\left[-\tilde{W}_{2}+W_{2}+\frac{1}{2}\tilde{W}_{1}'+\tilde{W}_{1}\partial\right]\delta (x-y)
\label{AII5a}\\
&&\{\tilde{J}_{3/2},J_{5/2}\}=\left[\tilde{W}_{3}-W_{3}-\tilde{W}_{2}'-\frac{1}{2}W_{2}'
-\tilde{W}_{2}\tilde{W}_{1}+2W_{2}\tilde{W}_{1}\right.\nonumber\\
&&\left.+\frac{3}{2}\tilde{W}_{1}'\tilde{W}_{1}+\frac{1}{3}\tilde{W}_{1}''
-\left(\tilde{W}_{2}+W_{2}-\frac{1}{2}\tilde{W}_{1}'-\tilde{W}^{2}_{1}\right)\partial\right]\delta (x-y)
\label{AII5b}\\
&&\{\tilde{J}_{5/2},J_{3/2}\}=\left[-\tilde{W}_{3}+W_{3}+\frac{1}{2}W_{2}'
+\tilde{W}_{2}\tilde{W}_{1}-2W_{2}\tilde{W}_{1}\right.\nonumber\\
&&\left.-\frac{1}{2}\tilde{W}_{1}'\tilde{W}_{1}+\frac{1}{6}\tilde{W}_{1}''
+\left(\tilde{W}_{2}+W_{2}+\frac{1}{2}\tilde{W}_{1}'-\tilde{W}^{2}_{1}\right)\partial\right]\delta (x-y)
\label{AII5c}\\
&&\{\tilde{J}_{5/2},J_{5/2}\}=\left[\tilde{W}_{4}-W_{4}-\frac{1}{2}\tilde{W}_{3}'-W_{3}'
-\tilde{W}_{3}\tilde{W}_{1}+3W_{2}\tilde{W}_{1}\right.\nonumber\\
&&+W_{2}\tilde{W}_{2}+\frac{3}{2}W_{2}'\tilde{W}_{1}+\frac{1}{6}\tilde{W}_{2}'\tilde{W}_{1}
+\frac{3}{2}W_{2}\tilde{W}_{1}'+\tilde{W}_{2}\tilde{W}_{1}'-3W_{2}\tilde{W}^{2}_{1} \nonumber\\
&&+\tilde{W}_{2}\tilde{W}^{2}_{1}-\frac{3}{10}W_{2}''-\frac{1}{10}\tilde{W}_{2}''-\tilde{W}^{2}_{2}
+\frac{1}{12}\tilde{W}_{1}''\tilde{W}_{1}+\frac{3}{4}\tilde{W}^{2}_{1}  \nonumber\\
&&+\frac{1}{12}\tilde{W}_{1}'''-\frac{3}{2}\tilde{W}_{1}'\tilde{W}^{2}_{1}-J_{3/2}'\tilde{J}_{3/2} 
-\left(\tilde{W}_{3}+2W_{3}-3W_{2}\tilde{W}_{1}-2\tilde{W}_{2}\tilde{W}_{1}\right. \nonumber\\
&&\left.\left.-\frac{1}{6}\tilde{W}_{1}''+W_{2}'-\tilde{W}^{3}_{1}+J_{3/2}\tilde{J}_{3/2}\right)\partial-W_{2}\partial^{2}
\right]\delta (x-y)
\label{AII5d}
\end{eqnarray}
and so on.

\end{document}